\documentclass[pra,twocolumn,showpacs,groupedaddress,superscriptaddress,aps,10pt]{revtex4-1}
\usepackage{bm,graphicx,amsmath}
\usepackage{placeins}
\usepackage{amssymb}
\usepackage{amsmath}
\usepackage{epsfig}
\usepackage{amssymb}
\usepackage{color}
\usepackage[colorlinks,linkcolor=blue,citecolor=blue]{hyperref}
\usepackage{subfigure}
\begin{document}

\title{Blue-detuned optical ring trap for Bose-Einstein condensates based on\\ conical refraction}
\date{\today}

\author{A. Turpin}\email{Corresponding author: alejandro.turpin@uab.cat}
\affiliation{Departament de F\'isica, Universitat Aut\`onoma de Barcelona, Bellaterra, E-080193, Spain}
\author{J. Polo}
\affiliation{Departament de F\'isica, Universitat Aut\`onoma de Barcelona, Bellaterra, E-080193, Spain}
\author{Yu. V. Loiko}
\affiliation{Aston Institute of Photonic Technologies, School of Engineering \& Applied Science Aston University, Birmingham, B4 7ET, UK}
\author{J. K\"uber}
\affiliation{Institut f\"ur Angewandte Physik, Technische Universit\"at Darmstadt, Schlossgartenstrasse 7, D-64289 Darmstadt, Germany}
\author{F. Schmaltz}
\affiliation{Institut f\"ur Angewandte Physik, Technische Universit\"at Darmstadt, Schlossgartenstrasse 7, D-64289 Darmstadt, Germany}
\author{T. K. Kalkandjiev}
\affiliation{Departament de F\'isica, Universitat Aut\`onoma de Barcelona, Bellaterra, E-080193, Spain}
\affiliation{Conerefringent Optics SL, Avda Cubelles 28, Vilanova i la Geltr\'u, E-08800, Spain}
\author{V. Ahufinger}
\affiliation{Departament de F\'isica, Universitat Aut\`onoma de Barcelona, Bellaterra, E-080193, Spain}
\author{G. Birkl}
\affiliation{Institut f\"ur Angewandte Physik, Technische Universit\"at Darmstadt, Schlossgartenstrasse 7, D-64289 Darmstadt, Germany}
\author{J. Mompart}
\affiliation{Departament de F\'isica, Universitat Aut\`onoma de Barcelona, Bellaterra, E-080193, Spain}

\begin{abstract}
We present a novel approach for the optical manipulation of neutral atoms in annular light structures produced by the phenomenon of conical refraction occurring in biaxial optical crystals. For a beam focused to a plane behind the crystal, the focal plane exhibits two concentric bright rings enclosing a ring of null intensity called the Poggendorff ring. We demonstrate both theoretically and experimentally that the Poggendorff dark ring of conical refraction is confined in three dimensions by regions of higher intensity. We derive the positions of the confining intensity maxima and minima and discuss the application of the Poggendorff ring for trapping ultra-cold atoms using the repulsive dipole force of blue-detuned light. We give analytical expressions for the trapping frequencies and potential depths along both the radial and the axial directions. Finally, we present realistic numerical simulations of the dynamics of a $^{87}$Rb Bose-Einstein condensate trapped inside the Poggendorff ring which are in good agreement with corresponding experimental results.
\end{abstract}

\date{\today }
%% 260.1960 Diffraction theory [IN: 260.0260 Physical optics]
%
 % activate for two-column option
%
\maketitle
\section{Introduction}
%(check bibliography of CR/comment bright rings for trapping or even studying)

Optical ring potentials (ORPs) with axial symmetry are considered as basic building blocks and the simplest nontrivial closed-loop circuits in atomtronics \cite{ATOMICS1,ATOMICS2,ATOMICS3,ATOMICS4,ATOMICS5} and atom interferometry \cite{2009_Cronin_RMP_81_1051}. Atoms can be trapped by means of the optical dipolar force in high or low intensity regions with red-detuned \cite{1998_Stamper-Kurn_PRL_80_2027, 2001_Barrett_PRL_87_010404} or blue-detuned \cite{1999_Ozeri_PRA_59_R1750} light, in what follows called bright and dark potentials, respectively. On the one hand, bright ORPs have been proposed and demonstrated with high-azimuthal-order Laguerre--Gaussian (LG) beams \cite{2000_Wright_PRA_63_013608} and also with annular microlenses \cite{2001_Birkl_OC_191_67}. Azimuthal lattices within ORPs have been demonstrated with time orbiting of light beams \cite{2008_Schnelle_OE_16_1405, 2008_Houston_JPB_41_211001} and by interference of LG beams of different azimuthal orders \cite{2007_FrankeArnold_OE_15_8619}. A one-dimensional stack of ORPs in a line has been proposed in an optical cavity \cite{2002_Freegarde_OC_201_99} and demonstrated with axicon beams \cite{2006_Courtade_PRA_74_031403}. Experimental storage and propagation of ultra-cold atoms and Bose-Einstein condensates (BECs) in bright ORPs have been reported recently \cite{2007_Ryu_PRL_99_260401, 2011_Ramanathan_PRL_106_130401}.
Dark ORPs on the other hand are optical fields with an annular region of minimum intensity \cite{2007_Olson_PRA_76_061404}, such as closed-loop optical singularities \cite{2001_Berry_PRSA_457_2251, 2010_Dennis_NatPhys_6_118}, for which the region of minimum intensity is exactly zero. For ultra-cold atoms, dark ORPs have the advantage of substantially reducing atom heating and decoherence rates \cite{1999_Ozeri_PRA_59_R1750} because of the low rate of spontaneous photon scattering as well as producing intrinsically flat potential minima. 
Blue-detuned ORPs have been experimentally reported by means of LG beams generated with spatial light modulators (SLMs) \cite{kivshar2013} and by amplitude masks \cite{masks1,masks2,masks3}. These two techniques might experience the following limitations: (i) a significant fraction of the input power is lost and, therefore, it does not contribute to create the optical trap, (ii) the smoothness and, therefore, the quality of the trapping potential is limited by the size and number of pixels for the SLMs and the resolution of the printing system for the amplitude masks, and (iii) an accurate control on the position and alignment of the optical elements being used is required. As a consequence, these two techniques yield typically not null intensity minima.
Producing ORPs with zero-intensity annular regions both along the radial and axial directions is a challenging task. In this case, the dark potential forms toroidal dark focus, i.e., a region of minimum intensity confined by higher intensities (light walls) both in the axial and radial directions. A toroidal dark focus has only been demonstrated using a superposition of two LG beams \cite{2007_Olson_PRA_76_061404}.

In this article, we present a new method to generate a dark ORP by means of the phenomenon of conical refraction (CR) \cite{hamilton,lloyd,2007_Berry_PO_55_13,1978_Belskii_OS_44_436,1999_Belsky_OC_167_1,2000_Belafhal,2004_Berry_JOA_6_289,2008_Kalkandjiev_SPIE_6994}, occurring in biaxial crystals. CR leads to a set of two concentric bright rings enclosing a dark ring of null intensity, known as Poggendorff dark ring (PDR). 
We theoretically investigate the three-dimensional (3D) field distribution around the CR PDR and show both theoretically and experimentally that it is a toroidal dark focus. We also discuss the use of the PDR as a blue-detuned ORP for ultra-cold atoms and demonstrate this with a $^{87}$Rb BEC.
This technique has the advantage of the easy generation of the toroidal dark trap, which only needs a focused Gaussian beam and a biaxial crystal. In addition, CR provides the full conversion of the input power into the toroidal dark trap, in contrast to the previously reported methods which introduce losses due to diffraction in the generation of LG beams. These features make the CR toroidal dark-focus beam very attractive for particle \cite{vault} and atom \cite{2000_Wright_PRA_63_013608, 2001_Birkl_OC_191_67, 2007_Olson_PRA_76_061404} trapping, in particular with blue-detuned light beams \cite{1999_Ozeri_PRA_59_R1750}. 

The article is organized as follows: In Section \ref{secCR} we give an introduction to the CR phenomenon, presenting its main fundamental characteristics: Section \ref{secPCR} presents the exact paraxial solution of the light pattern after propagation along one of the optic axes of a biaxial crystal, while its asymptotic approximation is presented in Section \ref {secACR}. In Section \ref{secatomsCR}, we investigate the use of the toroidal dark trap provided by CR for the trapping of ultra-cold atoms with blue-detuned light. In Section \ref{harmonic_approach} we apply the harmonic approximation around the PDR and present expressions for the trapping frequencies and height of the potential barriers as a function of the physical parameters of the trapping system. Then, both numerical simulations and experimental data for a $^{87}$Rb BEC trapped in the PDR are shown in Section \ref{secNUM}. Finally, the main conclusions are summarized	 in Section \ref{secCON}. 

\section{Conical refraction}
\label{secCR}
\begin{figure}[ht!]
\centering
\includegraphics[width=0.7 \columnwidth]{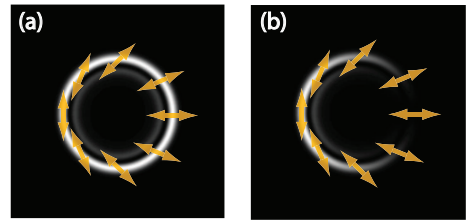}
\caption{Intensity and polarization distribution (depicted with yellow double arrows) of conical refraction with input beams of circular (a) and linear vertical (b) polarization. The dark ring between the two bright ones is known as the Poggendorff dark ring (PDR).}
\label{figCR_cone_ring}
\end{figure}

Conical refraction \cite{2007_Berry_PO_55_13, 1978_Belskii_OS_44_436, 1999_Belsky_OC_167_1,2000_Belafhal, 2004_Berry_JOA_6_289, 2008_Kalkandjiev_SPIE_6994} is observed when a focused light beam passing along an optic axis of a biaxial crystal (BC) is transformed into a light ring at the focal plane. Each pair of diagonally opposite points of the CR ring are orthogonally linearly polarized, as shown by double arrows in Fig.~\ref{figCR_cone_ring}. Therefore, a complete ring with uniform azimuthal intensity distribution is observed only for randomly (RP) or circularly (CP) polarized input beams (Fig.~\ref{figCR_cone_ring}(a)), while for linearly polarized (LP) input beams the intensity pattern is azimuthally crescent with a zero-intensity point, where the polarization is orthogonal to the one of the input beam. This polarization distribution, which constitutes an essential signature of the CR phenomenon, significantly differs from the usually known radial and azimuthal polarization modes and only depends on the orientation of the plane of the crystal optic axes \cite{1999_Belsky_OC_167_1}. 

The CR geometric optical approximation of the ring radius, $R_0$, is the product of the crystal length, $l$, and the CR semi-angle $\alpha$, i.e., $R_0=l \alpha$ \cite{2008_Kalkandjiev_SPIE_6994}. The CR semi-angle $\alpha$ depends on the principal refractive indices of the crystal as ${\alpha = \sqrt{(n_2^2-n_1^2)(n_3^2-n_2^2)/n_2^2}}$, where we have assumed $n_{1}<n_{2}<n_{3}$.
Moreover, under conditions of $\rho_0 \equiv \frac{R_0}{w_0} \gg 1$ the CR pattern at the focal plane is formed by a pair of concentric bright rings separated by the Poggendorff dark ring (Fig.~\ref{figCR_cone_ring}). Here, $w_0$ is the waist of the focused input beam, defined as the radius of the $e^{-2}$ relative intensity, i.e. $I(r=w_0) = e^{-2} I(r=0)$. 
Finally, as far as CR beam evolution is concerned, the focal plane is a symmetry plane along the beam propagation direction \cite{2008_Kalkandjiev_SPIE_6994}. The CR rings are observed at the focal plane, and more involved structures including secondary rings are found as one moves along the beam propagation direction. At points given by $z_{\rm{Raman}} = \pm \sqrt{4/3} \rho_0 z_R$ from the focal plane a bright spot known as the Raman spot \cite{vault} appears on the beam axis, where $z_R$ denotes the Rayleigh range of the focused input beam. In this section we describe the properties of the optical field at and close to the dark region of the Poggendorff ring. 

\subsection{Paraxial solution of the intensity distribution for CR}
\label{secPCR}
The paraxial solution describing CR was derived by Belsky and Khapalyuk \cite{1978_Belskii_OS_44_436} and later reformulated by Berry \cite{2007_Berry_PO_55_13}. For a uniformly polarized and cylindrically symmetric input beam it gives an electric field vector:
\begin{equation}
\vec{E} \left( \vec{\rho},Z\right)=
\left(\begin{array}{cc}
B_{C} + C & S \\ 
S & B_{C} - C
\end{array}
\right)
\vec{e}_{\rm{0}}
~,\quad \label{Eqs_output_beam_uniform} \\
\end{equation}
where $C=B_{S} \cos \left( \varphi + \varphi_0 \right)$ and $S=B_{S} \sin \left( \varphi + \varphi_0 \right)$. $\varphi$ is the azimuthal component in cylindrical coordinates and $\varphi_0$ is the orientation of the plane of the optic axes of the crystal. $E_{\rm{0}}$ and $\vec{e}_{\rm{0}}= \left( e_{\rm{x}}, e_{\rm{y}} \right)$ are the amplitude and unit vector of the electric field $\vec{E}_{\rm{0}}=E_{\rm{0}}\left(\rho , z \right)\vec{e}_{\rm{0}}$ of a focused input beam with waist $w_{0}$ and Rayleigh range $z_{R}$.  $Z \equiv z / z_{R}$ and $\vec{\rho} \equiv \left( \cos \varphi,\sin \varphi \right) r /w_{0}$ with $ \rho = \left| \vec{\rho} \right| \equiv r/w_0$ define, respectively, normalized axial and radial components in cylindrical coordinates with origin at the ring center ($\rho=0$) at the focal plane ($Z=0$).
$B_{C}$ and $B_{S}$ are the main integrals of the Belsky--Khapalyuk--Berry (BKB) solution, which describes the general properties of the CR beam. These integrals read:
\begin{eqnarray}
&B_{C}&(\rho,Z)=\frac{1}{2\pi}\int^{\infty }_{0} \eta a \left( \eta \right) e^{-i \frac{Z}{4} \eta ^{2} } \cos \left( \eta \rho_{0}\right) J_{0}\left( \eta \rho \right) d\eta, \quad
\\&B_{S}&(\rho,Z) = \frac{1}{2\pi}{\int}^{\infty }_{0} \eta a \left( \eta \right) e^{-i \frac{Z}{4} \eta ^{2} } \sin \left( \eta \rho_{0}\right) J_{1}\left( \eta \rho \right) d \eta, \;\;\quad
\end{eqnarray}
where $\eta \equiv \kappa w_0$, $\kappa$ being the spatial wave-vector and $J_{q}$ is the $q^{th}$-order Bessel function of the first type and $a \left( \eta \right)=2\pi\int_{0}^{\infty} r E_{\rm{0}} \left( r \right) J_{0} \left( \eta r \right) d r$ is the radial part of the 2D transverse Fourier transform of the input beam. 
For CP and LP inputs the intensity distribution behind the crystal becomes, respectively
\begin{eqnarray}
I_{\rm{CP}} &=& \left| B_{C} \right|^{2} + \left| B_{S} \right|^{2} ~,\quad 
\label{Eqs_output_beam_intensity_CP} \\
I_{\rm{LP}} &=& I_{\rm{CP}} + 2  Re \left[ B_{C} B_{S}^{*} \right] \cos \left( 2 \Phi - \left( \varphi + \varphi_0 \right) \right),~ 
\label{Eqs_output_beam_intensity_LP}
\end{eqnarray}
where $\Phi$ is the polarization azimuth of the linearly polarized input light with $\vec{e}_{\rm{0}} = \left( \cos \Phi , \sin \Phi \right)$.

\subsection{Asymptotic solution close to the Poggendorff dark ring}
\label{secACR}
The asymptotic solution for the Poggendorff dark ring, i.e., for $\rho_0 = \frac{R_{0}}{w_{0}} \gg 1$, is obtained by using the asymptotic expansion of Bessel functions $\cos \left( \eta \rho_{0}\right) J_{0}\left( \eta \rho \right) \approx \sin \left( \eta \rho_{0}\right) J_{1}\left( \eta \rho \right) \approx \cos{\left( \eta \xi - \pi /4 \right)} / \sqrt{2 \pi \eta \rho_{0}}$. Here we have centered the normalized radial component in cylindrical coordinates at $\rho_0$ by using $\xi \equiv \rho-\rho_{0} = r /w_0 - R_0/w_0$. In this case $B_{C} \approx B_{S}$ and the electric field can be written as \cite{1978_Belskii_OS_44_436, 1999_Belsky_OC_167_1}:
\begin{eqnarray}
\vec{E}\left( \xi, Z, \varphi \right) &&=  f \left( \xi , Z \right) E_{\rm{0}} \left( \vec{e}_{\rm{CR}} \cdot \vec{e}_{\rm{0}} \right) \vec{e}_{\rm{CR}}  ~,
\label{Eqs_output_beam_uniform_factored} 
\end{eqnarray}
where
\begin{eqnarray}
f \left( \xi , Z \right) =\sqrt{\frac{1}{8 \pi^3 \rho_0}} &&\int^{\infty }_{0}  d\eta
\sqrt{\eta} a\left(\eta \right)
e^{-i\frac{Z}{4} \eta ^{2}} \cos \left( \eta \xi - \frac{\pi}{4} \right),\nonumber\\
\label{Eqs_CR_Ring_Polarization}
\end{eqnarray}
and
\begin{eqnarray}
\vec{e}_{\rm{CR}}&&=
\left(
\begin{array}{r}
\cos \frac{\varphi + \varphi_0}{2} \\
\sin \frac{\varphi + \varphi_0}{2} 
\end{array}
\right) ~.
\label{Eqs_fxi_general}
\end{eqnarray}
Therefore, the asymptotic intensity distributions $I_{\rm{CP}}^{\rm{a}}$ and $I_{\rm{LP}}^{\rm{a}}$ for CP and LP input beams are, respectively,
\begin{eqnarray}
I_{\rm{CP}}^{\rm{a}} \left( \xi, Z \right)&=& \left| f \left( \xi , Z \right) \right|^{2}
~,\quad
\label{Eqs_CR_Ring_Intensity_CP} \\
I_{LP}^{\rm{a}} \left( \xi, Z, \varphi \right)&=& I_{\rm{CP}}^{\rm{a}} \cos^{2} \left( \Phi - \frac{\varphi + \varphi_0}{2} \right)
~.\quad
\label{Eqs_CR_Ring_Intensity_LP}
\end{eqnarray}
For LP input beams (see Eq.~(\ref{Eqs_CR_Ring_Intensity_LP})), the output intensity distribution lacks azimuthal symmetry. In this case the CR ring has a maximum and a zero intensity at azimuthal angles ${\varphi_{max} = 2 \Phi - \varphi_0}$ and ${\varphi_{min}= \varphi_{max} + \pi}$, respectively. These points possess, correspondingly, the same and the orthogonal polarization relative to that of the input beam, respectively (Fig.~\ref{figCR_cone_ring}(b)). 

In the following we will analyze the case of a CP input beam, for which the CR output intensity is azimuthally symmetric and its spatial distribution is described by Eq.~(\ref{Eqs_CR_Ring_Intensity_CP}). For a Gaussian input beam with normalized transverse profile of the electric field amplitude ${E(\rho)=\sqrt{2 P/\pi w_{0}^{2}} \exp (-\rho^2)}$, its Fourier transform is ${ a \left( \eta \right)=\sqrt{\frac{2 \pi P}{w_0^2}}\exp \left( -\eta^{2}/ 4  \right)}$. $P$ is the power of the input beam. For this case, Eq.~(\ref{Eqs_CR_Ring_Polarization}) can be analytically evaluated through the Kummer confluent hyper-geometric function $_{1}F_{1}(a;b;z)$ \cite{2004_Berry_JOA_6_289}:
\begin{eqnarray}
f \left( \xi , Z \right) &=&
\frac{\sqrt{P}}{(w_Z)^{3/4}\sqrt{2 \pi^2 w_0^2 \rho_0}}\left[\Gamma\left(\frac{3}{4}\right) \,_1F_{1}\left(\frac{3}{4};\frac{1}{2};-\frac{\xi^2}{w_Z}\right)\right.\nonumber \\
&+&\left. 2\frac{\xi}{\sqrt{w_{Z}}}\Gamma\left(\frac{5}{4}\right) \,_1F_{1}\left(\frac{5}{4};\frac{3}{2};-\frac{\xi^2}{w_Z}\right)\right]
% &=&
,~
\label{Eqs_fxi_Gauss_general_BKS_WeberFns}
\end{eqnarray}
where $w_{Z} = 1 + i Z$.

\begin{figure}[ht!]
\centering
\includegraphics[width=1 \columnwidth]{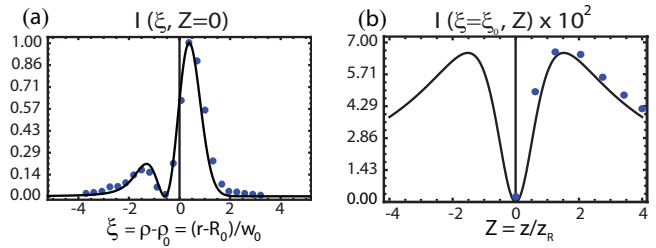}
\caption{Normalized CR intensity for a CP Gaussian input beam as given by Eq.~(\ref{Eqs_fxi_Gauss_general_BKS_WeberFns}) along the radial direction (a) at the focal plane and (b) along the axial direction at the radial position of the PDR ($\xi = \xi_0$). Blue solid circles represent experimental data with an experimental uncertainty of 5 $\%$ along both axis.}
\label{figPoggendorfDark_1D_X_Z}
\end{figure}

\begin{figure}[ht!]
\centering
\includegraphics[width=0.8 \columnwidth]{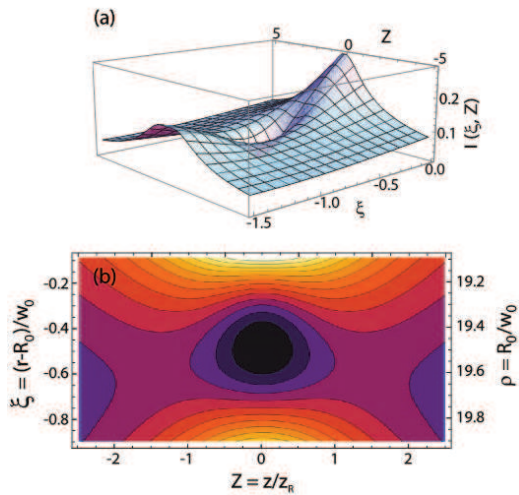}
\caption{(a) Normalized light intensity in three dimensions near the PDR. (b) 2D contour density plot near the PDR of the normalized light intensity calculated from Eqs.~(\ref{Eqs_CR_Ring_Intensity_CP}) and (\ref{Eqs_fxi_Gauss_general_BKS_WeberFns}) and for $\rho_0=R_0/w_0=20$. Color map: black = null intensity, white = high intensity.}
\label{figPoggendorfDark_2D_X0_Z}
\end{figure}

The solid line in Fig.~\ref{figPoggendorfDark_1D_X_Z}(a) shows the square modulus of  Eq.~(\ref{Eqs_fxi_Gauss_general_BKS_WeberFns}) at the focal plane ($Z=0$). $f(\xi_0,0)~=~0$, gives the radial position of the Poggendorff dark ring at the focal plane, being $\xi_0=-0.541$. In other words, the radius of the PDR is smaller than the geometric approximation of the CR ring, $R_{0}$, by approximately half the waist of the input beam. Note that $\xi = \rho-\rho_{0}$, with $\rho_{0} \equiv R_0 / w_0$. In the radial direction the PDR is confined by two maxima at $\xi_+ = 0.390$ and $\xi_- = -1.235$, respectively (see Table~\ref{table1}). 
Along the Z direction, the lowest intensity barrier is observed also at the radial position of the PDR, i.e., at $\xi=\xi_{0}$ as shown in Fig.~\ref{figPoggendorfDark_1D_X_Z}(b). At this radial point the positions of the intensity maxima along $Z$ obtained from Eq.~(\ref{Eqs_CR_Ring_Intensity_CP}) and (\ref{Eqs_fxi_Gauss_general_BKS_WeberFns}) are $Z_{\pm} = \pm 1.519$. Therefore, the PDR is confined by walls of light in all directions and forms a toroidal dark-focus. 
Table \ref{table1} presents the positions of the PDR and of the maxima in the radial ($\xi_{\pm}$) and axial ($Z_{\pm}$) directions.
\begin{table}[ht!]
  \centering
  \caption{Positions of the Poggendorff dark ring and of the maxima in the radial ($\xi \pm$) and axial ($Z \pm$) directions.}
  \begin{tabular}{rcc} \\ \hline
    Point name 											& \quad$\xi$ 						&\quad $Z$ 				 \\ \hline
    Dark Ring: $\xi_0$ 	& \quad-0.541 					&\quad 0  						 \\ 
    Bright Rings: $\xi_+$ 	&\quad 0.390  &\quad 0 						  \\ 
                  $\xi_-$ 	&\quad-1.235  &\quad 0 						  \\ 
    Maxima along Z: $Z_{\pm}$ &\quad -0.541 					&\quad $\pm$1.519 	 
  \end{tabular}
  \label{table1}
\end{table}
As a visualization of the toroidal dark trap provided by the PDR of CR, Fig.~\ref{figPoggendorfDark_2D_X0_Z}(a) shows the three-dimensional distribution of light intensity of the asymptotic approximation of the BKB solution near the focal plane. Fig.~\ref{figPoggendorfDark_2D_X0_Z}(b) is a contour plot near the PDR, confirming that it is a region of low intensity surrounded in all directions by regions of higher intensity. 
Note that the PDR is an exact null intensity region only for input Gaussian beams under the asymptotic approximation, i.e. for $\rho_0 \gg 1$, while non-zero intensity radial minimum points are found out of the paraxial approximation, as reported in Refs. \cite{hole,SGCR}. For input beams with different transverse profile the CR pattern may change \cite{peet,dublin}.

We have experimentally checked that near the PDR the light intensity increases in all directions, see blue solid circles in Fig.~\ref{figPoggendorfDark_1D_X_Z}(a) and Fig.~\ref{figPoggendorfDark_1D_X_Z}(b). These experiments on the CR PDR were carried out using a CP focused input Gaussian beam ($w_0 = 40 \,\rm{\mu m}$, $z_R = 7.9\, \rm{mm}$) at $\lambda = 640\,\rm{nm}$ and a KGd(WO$_4$)$_2$ biaxial crystal (cross-section $6\times4\,\rm{mm}^2$, $l = 28\,\rm{mm}$, $\alpha = 17\, \rm{mrad}$) cut perpendicular to one of the optic axes (entrance surface parallelism better than $10$ arc seconds) yielding a CR ring of $R_{0} = 476\,\rm{\mu m}$ ($\rho_0 \approx 12$). The transverse light patterns at and around the focal plane were recorded with a CCD camera. 

\section{Application to atom trapping with blue-detuned light}
\label{secatomsCR}
\subsection{Harmonic potential approximation}
\label{harmonic_approach}
In the previous sections we have described the intensity distribution near the PDR in the radial and axial directions, showing that this region is a dark ORP. This makes the PDR a good candidate for atom trapping applications with blue-detuned light. For a given light intensity $I(\vec{r})$, the trapping dipole potential reads ${U(\vec{r})= \tilde{U}_0 I(\vec{r})}$, where for alkali atoms for sufficiently large detuning and linear polarization
\begin{eqnarray}
\tilde{U}_0 &=& -\frac{\pi c^2}{2} \left[ \frac{\Gamma_{D_2}}{\omega^3_{D_2}} \left( \frac{2}{\omega_{D_2}-\omega_L} \right ) + \frac{\Gamma_{D_1}}{\omega^3_{D_1}} \left( \frac{1}{\omega_{D_1}-\omega_L} \right ) \right ],\qquad 
\label{auxpot}
\end{eqnarray}
%
%
%\begin{align}
%U(\vec{r})= -  &\left[ \frac{\pi c^2 \Gamma_{D_2}}{2 \omega^3_{D_2}} \left( \frac{2}{\omega_{D_2}-\omega_L} \right )+\right.\notag \\
%&\left. \frac{\pi c^2 \Gamma_{D_1}}{2 \omega^3_{D_1}} \left( \frac{1}{\omega_{D_1}-\omega_L} \right ) \right ] I(\vec{r}),
%\label{potential}
%\end{align}
%
as given e.g. in {\cite{grimm2000}}.
Note that in $\tilde{U}_0$ we have applied the rotating-wave approximation. Here, $c$ is the speed of light in vacuum, $\Gamma_{D_{i}}$ and $\omega_{D_{i}}$ ($i=1,2$) are, respectively, the natural line width and frequency of the $D_i$ line of the atomic species, and $\omega_L$ is the frequency of the input beam. In our case, $I(\vec{r})$ is given by Eq.~(\ref{Eqs_CR_Ring_Intensity_CP}) and Eq.~(\ref{Eqs_fxi_Gauss_general_BKS_WeberFns}) and $\tilde{U}_0 > 0$ for blue-detuned light. By using the harmonic oscillator approximation, we have obtained the following expressions for the corresponding radial ($\omega_r$) and axial ($\omega_z$) trapping frequencies of the PDR ($\xi = \xi_0, Z=0$)
\begin{eqnarray}
\omega_{r,z} &=& \sqrt{ \frac{A_{r,z} \tilde{U}_{0} P}{\pi^2 m w^4_0 \rho_0}}~,
\label{omegar}
\end{eqnarray}
with $m$ being the atomic mass and the numerical constants $A_{r}(Z=0) = 4.63$ and $A_{z} = 0.34$. Eq.~(\ref{omegar}) is obtained by expanding Eq.~(\ref{Eqs_fxi_Gauss_general_BKS_WeberFns}) in Taylor series, introducing it into Eq.~(\ref{Eqs_CR_Ring_Intensity_CP}) and considering the $\xi^2$ coefficient.
%Fig.\ref{potentialsharm} shows graphically the agreement between the original potential and its harmonic approximation. 
%
%\begin{figure}[ht!]
%\centering
%\includegraphics[width=1 \columnwidth]{f/potentials_harmonic.eps}
%\caption{Light potential near the PDR along the radial (a) and longitudinal (b) directions using Eqs.(\ref{Eqs_fxi_Gauss_general_BKS_WeberFns}) and (\ref{Eqs_CR_Ring_Intensity_CP}) (black solid lines) and its harmonic approximation (red dashed lines).}
%\label{potentialsharm}
%\end{figure}
%

Note that from Eqs.~(\ref{Eqs_CR_Ring_Intensity_CP}), (\ref{Eqs_fxi_Gauss_general_BKS_WeberFns})--(\ref{omegar}) for a given biaxial crystal, i.e. for a fixed $R_0$ and $w_{0}$, the trapping frequencies and the maxima of the potential barriers can be tuned by modifying the power $P$ and the frequency $\omega_L$ of the input beam. We have obtained that at the focal plane the maxima of the potential barriers are described by
\begin{equation}
U(\xi_\pm,0) = C_{\pm} \tilde{U}_0  \frac{P}{4 \pi^2 w_0^2 \rho_0},
\label{pot_inner}
\end{equation}
where $C_{+}=2.54$ (outer bright ring) and ${C_{-}=0.541}$ (inner bright ring).

\begin{figure}[ht!]
\centering
\includegraphics[width=1 \columnwidth]{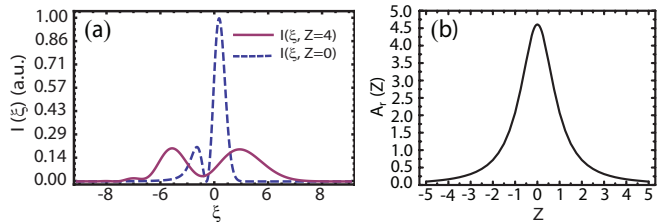}
\caption{ (a) Profile of the trapping potential at $Z=0$, i.e. at the focal plane (dashed curve), and at $Z=4$ (solid curve) where the inner and the outer bright rings of CR have equal maximum intensity. (b) Coefficient $A_{r}$ as a function of $Z$. The analytical expression for the $A_r(Z)$ is given by Eq.~(\ref{eqtrapz}).}
\label{figz4}
\end{figure}

There can be other experimental situations however, where it is required to work outside the focal plane, for instance in experiments where a more symmetric potential is needed, such as the one shown with a solid line in Fig.~\ref{figz4}(a), where the radial intensity distribution close to the PDR is shown for the focal plane ($Z=0$) and the plane $Z=4$. In these cases, Eq.~(\ref{omegar}) can be utilized to calculate the trapping frequency of the potential at any axial position $Z$ by just replacing $A_{r}(Z=0)$ by
\begin{equation}
A_r(Z) = -0.051+\frac{8.817}{1.873  +2.307 Z^2}.
\label{eqtrapz}
\end{equation}
Figure~\ref{figz4}(b) presents the dependence of $A_{r}(Z)$ on $Z$. 
Note that outside the focal plane an offset to the potential is occurring, since the minimum intensity point is no longer of null intensity as plotted for $Z=4$ as solid line in Fig.~\ref{figz4}(a). We have found that this non-zero minimum intensity point
%, which can increase the scattering rate of atoms, 
can be taken into account by means of the optical potential along the axial direction, 
\begin{equation}
U(\xi_0,Z) = \tilde{U}_{0} \frac{P}{4 \pi^{2} w^2_0 \rho_0}Z^2.
\label{potz}
\end{equation}
The confining maxima along $Z$ are not well described by the harmonic approximation and must be evaluated using Eqs.~(\ref{Eqs_CR_Ring_Intensity_CP}) and (\ref{Eqs_fxi_Gauss_general_BKS_WeberFns}). They read
\begin{equation}
U(\xi_0,Z_{\pm}) = 0.17 \tilde{U}_0  \frac{P}{4 \pi^2 w_0^2 \rho_0}.
\label{pot_z}
\end{equation}

\subsection{Numerical simulations of a BEC of $^{87}$Rb atoms}
\label{secNUM}
To demonstrate the applicability of the PDR for ultra-cold gases, now we discuss the two-dimensional (2D) evolution of a BEC of $^{87}$Rb atoms confined in an annular geometry within the focal plane by using the PDR of CR and a strong additional confinement along the axial direction so that $\omega_{\rm{axial}} \gg \omega_r$ \cite{birkl2013}. Such confinement can be achieved by using an additional red-detuned sheet of light (e.g. generated by focusing a Gaussian beam with a cylindrical lens) to compensate for the weak axial confinement as well as, in case of a horizontal ring plane, the effect of gravity as shown in \cite{birkl2013}. We use the 2D Gross--Pitaevskii equation (GPE) in order to study the dynamics of the BEC along the ring potential:
\begin{equation}
i\hbar\frac{\partial}{\partial t}\Psi(\vec{r},t)=\left(-\frac{\hbar^2}{2m}\vec{\nabla}^2+V_{\rm{ext}}(\vec{r})+g_{2D}\left|\Psi(\vec{r},t)\right|^{2}\right)\Psi(\vec{r},t),
\end{equation}
where $V_{\rm{ext}}(\vec{r})$ is the external potential, ${g_{2D}=2\hbar a_{s}N\sqrt{2\pi\frac{\hbar \omega_{z}}{m}}}$, $a_{s}$ is the scattering length, $\omega_{z}$ is the frequency of the confining potential in the axial direction, $m$ is the mass of the $^{87}$Rb atoms and $N$ is the number of atoms.

In our simulations, we consider trapping close to the $D_2$ and $D_1$ lines of $^{87}$Rb. These lines possess natural line widths of ${\Gamma_{D_2}=2 \pi \times 6.07\,\rm{MHz}}$ and ${\Gamma_{D_1}=2 \pi \times 5.75\,\rm{MHz}}$ and frequencies of $\omega_{D_2} = 2 \pi \times 384.23\,\rm{THz}$ and ${\omega_{D_1} =2 \pi \times 377.11\,\rm{THz}}$, respectively. Thus, to calculate the trapping frequencies and the maxima of the potential barriers is straightforward by using Eqs.~(\ref{auxpot}), (\ref{omegar}), (\ref{pot_inner}), and (\ref{pot_z}). Based on the experimental parameters of \cite{birkl2013}, for a biaxial crystal yielding a CR ring of $R_0=170\,\rm{\mu m}$, an input beam waist $w_0=18\,\rm{\mu m}$, a light frequency of $\omega_L = 2 \pi \times 378.40\,\rm{THz}$ and a laser power $P=27\,\rm{mW}$, at the focal plane, the maxima of the potential barriers and trapping frequencies are, respectively, $U(\xi_-,Z=0)/k_B = 280\,\rm{nK}$, $U(\xi_+,Z=0)/k_B = 1314\,\rm{nK}$ and $\omega_r = 2\pi\times265\,\rm{Hz}$, where $k_B$ is the Boltzmann constant. 

\begin{figure*}[hbt!]
\centering
\includegraphics[width=1.8 \columnwidth]{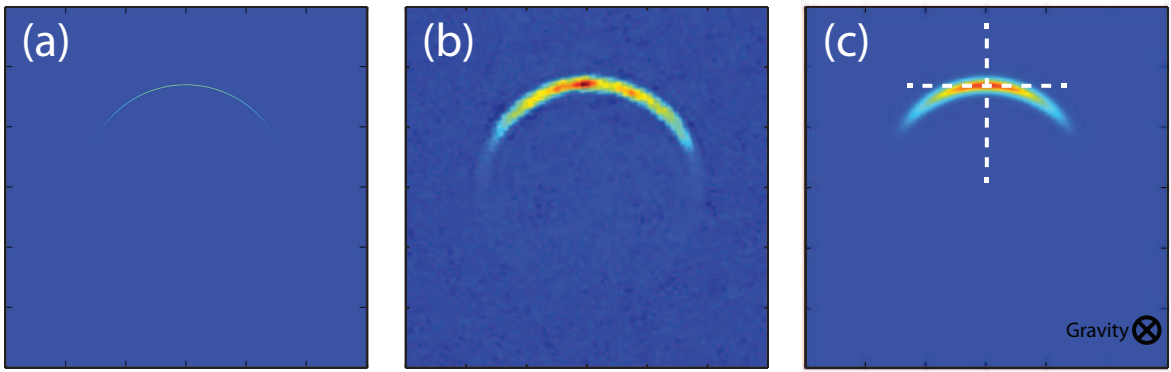}
\caption{(a) Plot of the atomic density from the numerical simulation of a trapped $^{87}$Rb BEC after $30\,\rm{ms}$ of expansion in the ring $V_{r}=\frac{1}{2}m\omega_{r}^{2}\left(r-(R_{0}-0.541 w_0) \right)^{2}$, with the frequency $\omega_{r}=2\pi\times 265 \,\rm{Hz}$ being calculated using the harmonic approximation. Parameter values used for the simulation: $R_0=170\,\rm{\mu m}$, $w_0=18\,\rm{\mu m}$, $P=27\,\rm{mW}$, $w_{z}=2\pi\times500\,\rm{Hz}$, $a_s=5.45\,\rm{nm}$ and $N=12000$ atoms. 
(b) Experimental density distribution of a trapped $^{87}$Rb BEC in the CR ring potential using the same experimental parameters as for the numerical simulation, with the exception of the axial confinement, that was made using a red-detuned Gaussian beam focused with a cylindrical lens, providing a measured trapping frequency of $w_{z}^{\rm{exp}}=2\pi\times (169 \pm 2) \,\rm{Hz}$. The measured radial trapping frequency provided by the CR PDR was $\omega_{r}^{\rm{exp}}=2\pi\times (300 \pm 20) \,\rm{Hz}$.
(c) Numerical simulation under the same conditions as in (a) but including the scattering induced by the position spreading during detection.
Each Fig. is $600\,\rm{\mu m} \times 600\,\rm{\mu m}$. Color map: dark blue (red) corresponds to null (high) intensity. White dashed lines in (c) indicate the position of the cross-dipole trap with respect to the PDR, being both of them orthogonal to the gravity field. The waist radius of each beam from the cross-dipole trap is $25\,\rm{\mu m}$.}
\label{figexp}
\end{figure*}
Figure~\ref{figexp}(a) shows the numerical simulation for a $^{87}$Rb BEC, with scattering length $a_s=5.45\,\rm{nm}$ , of $N=12000$ atoms trapped in a blue-detuned harmonic annular potential $V_{r}=\frac{1}{2}m\omega_{r}^{2}\left(r-(R_{0}-0.541 w_0)\right)^{2}$ with radial frequency $\omega_{r}=2\pi\times 265\,\rm{Hz}$ calculated using Eq.~(\ref{omegar}). Our numerical simulations are based in the following loading process: the BEC is created in a cross-dipole trap, see e.g. \cite{birkl2011}, and loaded into the red-detuned sheet of light.  
We consider that both the cross-dipole trap and the red-detuned sheet of light are orthogonal to the gravity field. The PDR potential, which also lies orthogonal to the gravity field, is placed tangent to one of the beams of the cross-dipole trap, see Fig.~\ref{figexp}(c). The beam from the cross-dipole trap that is tangent to the PDR is switched off as the CR PDR potential is switched on, in an adiabatic process. 
Finally, the remaining beam from the cross-dipole trap is switched off and the BEC expands in the CR PDR potential. We plot the atomic density of the BEC after $30\,\rm{ms}$ of expansion in the annular potential. In order to reduce the transverse excitations, the loading of the BEC into the CR ring potential has been performed adiabatically (in our case during $20\,\rm{ms}$) as reported in \cite{birkl2013}. Fig.~\ref{figexp}(b) shows the corresponding experimental density distribution after $30\,\rm{ms}$ expansion of a $^{87}$Rb BEC trapped in the real CR PDR. The CR PDR was placed perpendicular to gravity and a sheet of light generated by focusing a Gaussian beam with a cylindrical lens was used to hold atoms against gravity. The corresponding measured trapping frequencies are $\omega_{r}^{\rm{exp}} = 2\pi\times (300 \pm 20) \,\rm{Hz}$ and $\omega_{z}^{\rm{exp}} = 2\pi\times (169 \pm 2) \,\rm{Hz}$. 
The major discrepancy between experimental and numerical density plots is found in the radial width of the BEC. In the ideal case (Fig.~\ref{figexp}(a)), the effects of broadening due to finite optical resolution and photon scattering of the detection light have not been considered to obtain the image, which shows a BEC with a width of $3\,\rm{\mu m}$. In contrast, the experimental image from Fig.~\ref{figexp}(b), which shows a BEC with a width of $25\,\rm{\mu m}$, was obtained by using red-detuned light ($\lambda_{\rm{ill}} = 780\,\rm{nm}$, $P_{\rm{ill}} = 0.25\,\rm{mW}$) to illuminate the BEC during a time of $t_{\rm{ill}} = 200\,\rm{\mu s}$. For this illuminating light we have calculated a scattering rate $\Gamma_{sc} = 3.29 \times 10^{6}\,\rm{s^{-1}}$ that, together with the recoil velocity of $v_{\rm{rec}} = 5.89\,\rm{mm/s}$, increases the width of the atomic cloud in the radial direction by $21.89\,\rm{\mu m}$ during the illumination time. 
Figure~\ref{figexp}(c) shows the same numerical simulation as Fig.~\ref{figexp}(a) where we have taken into account now the increase of the width produced by the detection process included. Now, numerical simulation and experimental result agree well. 
%
%The only disagreement between numerical simulation and experimental results is the with of the experimental trapped BEC, due to imperfections in the imaging and detection system (add thermal atoms and condensation fraction). 
%
\begin{figure}[ht!]
\centering
\includegraphics[width= 0.8 \columnwidth]{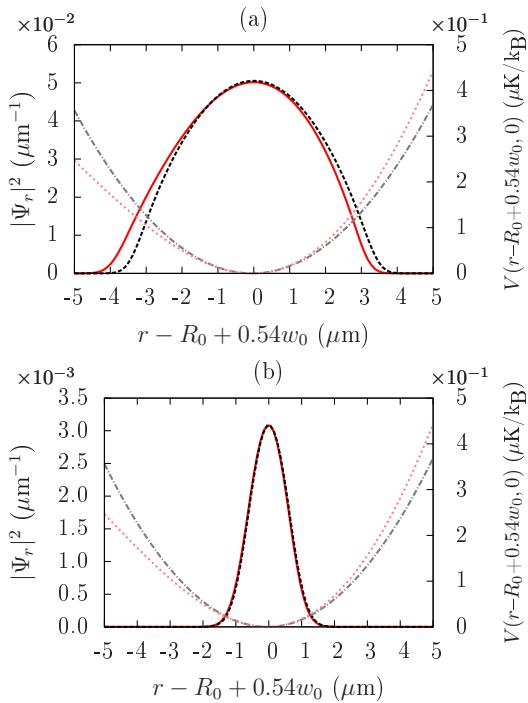}
\caption{Radial sections of the atomic density of the BEC (a) before and (b) after $30\,\rm{ms}$ of azimuthal expansion of the BEC trapped in the harmonic potential $V_{r}=\frac{1}{2}m\omega_{r}^{2}\left(r-(R_{0}-0.541 w_0) \right)^{2}$ (black-dashed line) and in the Poggendroff dark ring of CR (red-solid line). Black-dashed and red-dotted lines are the corresponding trapping potentials.
Parameter values: $R_0=170\,\rm{\mu m}$, $w_0=18\,\rm{\mu m}$, $P=27\,\rm{mW}$, $w_{r}=2\pi\times 265\,\rm{Hz}$, $w_{z}=2\pi\times500\,\rm{Hz}$, $a_s=5.45\,\rm{nm}$ and $N=12000$ atoms. The ground state (a) is obtained by adding an extra confinement ($w_{\rm{azi}}=2\pi\times 265\,\rm{Hz}$) in the azimuthal direction in order to reproduce the loading of the BEC in the CR ring trap.}
\label{fig_grounds}
\end{figure}

In order to further confirm the validity of the harmonic approximation, we also studied the ground state of the BEC trapped in the toroidal dark-focus (see Fig.~\ref{fig_grounds}). The physical system considered has the following parameters: $R_0=170\,\rm{\mu m}$, $w_0=18\,\rm{\mu m}$, $P=27\,\rm{mW}$, $w_{r}=2\pi\times 265\,\rm{Hz}$, $a_s=5.45\,\rm{nm}$ and $N=12000$ atoms. The toroidal dark trap is placed orthogonal to gravity and, therefore, to provide confinement along the axial direction we have considered a sheet of light analogous to the one discussed in \cite{birkl2013} with a trapping frequency $w_{z}=2\pi\times500\,\rm{Hz}$. 
The plots represent a section of the wave-function in the radial direction at the peak value of the density. 
The red-solid line in Fig.~\ref{fig_grounds}(a) shows the wave-function ground state of the BEC trapped in the PDR potential (represented by the red-dashed line), while the black-solid line is the ground state of the BEC trapped in the harmonically approximated potential (represented by black-dashed line) equivalent to the PDR. To provide confinement in the azimuthal direction, an extra beam yielding a trapping frequency of $w_{\rm{azi}} = 2 \pi \times 265\,\rm{Hz}$ is included. We have found a $0.7\%$ of relative difference between the energies of the two ground states.
Figure~\ref{fig_grounds}(b) presents the BEC wave-function after $30\,\rm{ms}$ of expansion within the harmonically approximated ring potential (black-solid line) and within the real PDR potential (red-solid line). Black- and red-dashed lines represent the harmonic ring potential and the PDR potential, respectively. In this case, the relative difference between both wave-functions is negligible. 
These results confirm the good agreement between the harmonic approximation derived in Section \ref{harmonic_approach} and the original PDR.

\section{Conclusions}
\label{secCON}

In summary, we have presented a novel approach for generating toroidal optical traps for ultra-cold neutral atoms by applying the PDR as a blue-detuned toroidal trap for BECs We have studied the normalized intensity distribution around the annular ring structure of the CR phenomenon in biaxial crystals. For a well developed CR ring, i.e. when $R_0 \gg w_0$, experimental results of the intensity distribution are compared with the exact paraxial solution and with its asymptotic approximation. 
We have found the positions of the bright and dark rings of CR and the position of the two points with maximum intensity along the beam propagation direction, both experimentally and analytically. We have shown that the radius of the PDR is smaller than the optical geometric approximation of the CR ring radius $R_{0}$, by approximately half the waist radius of the input beam ($-0.541 w_0$ in Table~\ref{table1}). All previous related works \cite{2007_Berry_PO_55_13, 1978_Belskii_OS_44_436, 1999_Belsky_OC_167_1,2000_Belafhal, 2004_Berry_JOA_6_289} were performed considering that the radius of the PDR exactly coincided with $R_0$. The reported results show that the PDR is enclosed by higher intensity walls both in the radial as in the axial directions, i.e. it is a toroidal dark-focus in all three dimensions, at variance with other light beams possessing only radial confinement, such as Laguerre--Gaussian modes.  
We have applied the harmonic approximation around the PDR and we have derived the expression for the radial and axial trapping frequencies and the maxima of the potential barriers for blue-detuned light as a function of common experimental parameters such as beam power, beam waist, detuning and the parameters of the crystal. 
The reported results show the suitability of the PDR for trapping ultra-cold atoms with blue-detuned light, making this technique ideal for experiments where well-defined potentials and high intensity beams are required \cite{2000_Wright_PRA_63_013608, 2001_Birkl_OC_191_67, 2007_Olson_PRA_76_061404, 1999_Ozeri_PRA_59_R1750,birkl2013}. 
Therefore, as a proof of the usefulness of the derived theory we have performed numerical simulations of the dynamics of a trapped $^{87}$Rb BEC with $N=12000$ atoms in the dark ring potential using the harmonic approximation and have compared the obtained results with the solution of the original CR light field. We have also compared the ground states in both cases and we have found $0.7\%$ relative difference in energy between them. The numerical simulations agree well with the experimental results on the dynamics of a trapped $^{87}$Rb BEC in the PDR of CR. 

The main advantages of the presented technique are the simple generation and high quality of the CR toroidal dark trap, since the only requirements are a biaxial crystal and a focused input Gaussian beam, at variance with the techniques using LG beams that need the interference of at least two beams \cite{2007_Olson_PRA_76_061404}. Also, the minimum (and practically null) intensity circle offered by the toroidal dark trap avoids photon scattering and presents no corrugation of the potential minimum at the focal plane.

Additionally, and at variance with techniques based on LG beams \cite{kivshar2013} or amplitude masks \cite{masks1,masks2,masks3}, the use of a biaxial crystal allows for the full input power to be converted into the CR dark toroidal trap. which increases the efficiency in ultra-cold atom trapping experiments. Moreover, biaxial crystals can be transparent to an extremely wide spectral range\cite{dublin2013} (0.35~$\rm{\mu m}$-5.5~$\rm{\mu m}$ in KGd(WO$_4$)$_2$, for instance),  in contrast with spatial light modulators used in the generation of LG beams, which only work in a small spectral range of few hundreds of $\rm{nm}$, typically.

A range of applications of this technique can be envisioned: for optimized beam geometries, i.e. small $w_0$, $R_0$, and $z_R$, the toroidal dark focus of the PDR generated by CR could be used to built an all-optical trap for BECs using a single beam. Under such conditions, this potential could be used as a basic element in atomic SQUID experiments \cite{SQUID,boshier1}, as well as to study the dynamics of matter waves with periodic boundary conditions and the generation of persistent currents \cite{2011_Ramanathan_PRL_106_130401,boshier2}. For large $R_0$, the PDR can be used as a dark 2D ring potential by using a 1D light sheet, along the axial direction, as an additional confining potential. This configuration would allow to study wave-packet interference in a mesoscopic ring simulating a quasi-one-dimensional system \cite{birkl2013}. By modifying this 1D light sheet to a blue-detuned double layer also accessible via CR \cite{ebs2013,hole} again a fully blue-detuned dark trap geometry with added flexibility is generated.

As an encouragement for future investigations, other cylindrically symmetric light structures of interest in atomic trapping experiments such as flat intensity regions or doughnut-like beams are also accessible via CR \cite{hole,SGCR}.
Additionally, a radial optical lattice could be generated by means of a cascade of biaxial crystals, generating $2^{N-1}$ dark rings for a cascade of N biaxial crystals \cite{berry2010, tur2013}. This could be also combined with the technique shown in \cite{2012_Turpin_OL_FSOC} to generate an azimuthal optical lattice with controllable number of nodes and separation between them, applicable in quantum-many body systems experiments \cite{ringlattice1,ahufinger_book}. 
Also interesting is the possibility of using the PDR to coherently injecting, extracting, and velocity filtering of particles, ultra-cold atoms and BECs as reported in \cite{menchon2014,vault} by tuning the polarization of the input beam and opening/closing the ring potential.
Finally, we would also like to note that by switching to red-detuned light, the inner and outer bright rings around the PDR generate an intrinsically concentric system of a double-ring potential which can be used for the generation of coherent double wave packets for the investigation of wave packet tunneling and coupled persistent currents of ultra cold atoms \cite{persistent_currents1,persistent_currents2}.

\section*{Acknowledgments}
The authors gratefully acknowledge financial support through the Spanish MICINN contract, and FIS2011-23719, the Catalan Government contract SGR2009-00347, and the DAAD contracts 50024895 and 57059126. A. Turpin acknowledges financial support through grant AP2010-2310 from the MICINN and the DAAD grant 91526836. J. Polo also acknowledges financial support from FPI Grant No. BES-2012-053447.

\end{document}